\documentclass[10pt]{article}  
\usepackage{graphicx}   
\usepackage[a4paper,left=3.0cm,right=3.0cm, bottom=3.0cm, top=3cm]{geometry}
\usepackage{apacite}
\usepackage{ragged2e}
\usepackage{amsmath}
\usepackage{gensymb}
\usepackage[capitalise]{cleveref}

\let\citeA\shortciteA
\let\cite\shortcite
\def\notation{\bgroup
\section*{Notation}
\description
\def\notation##1{\item[\boldmath ##1]}}
\def\endnotation{\enddescription\vskip12pt\egroup}

\usepackage{xcolor}

\usepackage{xurl}

\begin{document}

\begin{center}
\LARGE {\bf Depth Dependence of Coseismic Off-Fault Damage and its Effects on Rupture Dynamics\\[12pt]}
\large

Roxane Ferry$^{1}$, Marion Y. Thomas$^{2}$, Harsha S. Bhat$^{3}$, Pierpaolo Dubernet$^{3}$

\begin{enumerate}
\small
\setlength\itemsep{0.01em}
\item Institute of Civil Engineering, Institute of Materials Science and Engineering, École Polytechnique Fédérale de Lausanne (EPFL), Lausanne, Switzerland\\
\item Institut des Sciences de la Terre de Paris, Sorbonne Université, CNRS, UMR 7193, Paris, France\\
\item Laboratoire de Géologie, Département de Géosciences, École Normale Supérieure, CNRS, UMR 8538, PSL Université, Paris, France\\
\item[\large *] Corresponding Author: Roxane Ferry~~roxane.ferry@epfl.ch
\end{enumerate}
\end{center}

{\textit{\color{gray}{\small Manuscript submitted to JGR: Solid Earth \today}}}

\section*{Key Points:}
\begin{itemize}
\setlength\itemsep{1em}
\item We use a mechanical model to investigate how off-fault changes in material properties and generated waves affect earthquake slip dynamics
\item The off-fault damage zone narrows but becomes denser with depth, absorbing energy, stabilizing slip rates and reducing rupture velocity
\item Interactions between bulk changes and fault slip drive both SSE and earthquake dynamics, emphasizing the importance of the whole fault zone
\end{itemize}
\clearpage 

\section*{Abstract}
Faults are complex systems embedded in an evolving medium fractured by seismic ruptures. This off-fault damage zone is shown to be thermo-hydro-mechano-chemically coupled to the main fault plane by a growing number of studies. Yet, off-fault medium is still, for the most part, modelled as a purely elastic -- hence passive -- medium. Using a micromechanical model that accounts for dynamic changes of elastic moduli and inelastic strains related to crack growth, we investigate the depth variation of dynamically triggered off-fault damage and its counter-impact on earthquake slip dynamics. We show that the damage zone, while narrowing with depth, also becomes denser and  contrary to prevailing assumptions continues to act as an energy sink, significantly influencing rupture dynamics by stabilizing slip rates. Furthermore, we observe that damage formation markedly reduces rupture velocity and delays, or even prevents, the transition to supershear speeds even for a narrow damage zone. This underscores the critical need to incorporate the complex interplay between the main fault plane and its surrounding medium across the entire seismogenic zone. As a proof of concept, we introduce a 1D spring-slider model that captures bulk elastic variations, by modulating spring stiffness, and normal stress variations that emulate changes in bulk load. This simple model demonstrates the occurrence of slow slip events alongside conventional earthquakes, driven by the dynamic interaction between bulk temporal evolution and fault slip dynamics, without necessitating any changes to frictional properties.

\section*{Plain Language Summary}
Faults are part of a complex and evolving environment. Around these faults is a damaged rock area, known as the off-fault damage zone. Studies show this zone is interconnected with the main fault through thermal, hydraulic, mechanical, and chemical processes. However, most models treat the off-fault zone as purely elastic and passive medium. We use a model considering changes in material properties caused by cracking to understand how damage varies with depth and affects earthquake behavior. We find that the damage zone narrows but becomes denser with depth. Contrary to common beliefs, this zone continues to absorb energy, stabilizing the fault's motion during an earthquake. Our results show that damage generation slows down the rupture's speed and can prevent it from becoming supershear, even in narrow damage zones. This highlights the importance of considering interactions between the fault and its surrounding from the surface down to the end of the seismogenic zone. To extrapolate over the seismic cycle, we introduce a simple model simulating changes in surrounding material during fault slip. This model shows that both slow and fast earthquakes can occur due to dynamic interactions between the fault and the surrounding medium, without changing the fault's friction properties.

\clearpage

\section{Introduction}

\par Understanding on- and off-fault earthquake processes is crucial for effective mitigation of seismic hazard. To achieve this, comprehensive knowledge of fault zone physical properties and how it evolves through time and space is imperative. 

Observations reveal that a real fault deviates significantly from the simplified conception of a planar fault embedded in an elastic medium. Faults exhibit roughness across various scales and are more appropriately referred to as fault systems. These systems encompass geometrical complexities ranging from the kilometric regional fault structure down to the smaller-scale fracture networks. 
At smaller scale, a fault structure is composed of a fault core abutted by a damage zone --~a region characterized by intense fracturing due to fault slip~-- within the country rock. Based on field observations, damage density roughly decreases exponentially with distance from the fault \cite{mitchell_nature_2009} and induces changes in elastic properties 
\textcolor{black}{\cite{qiu_seismic_2021, pio_lucente_temporal_2010}}. Experimental evidences, such as the triaxial experiments conducted by \citeA{faulkner_slip_2006} and \citeA{mitchell_towards_2012}, demonstrates that Young's modulus decreases while Poisson's ratio and permeability rise with increasing damage.
Additionally, along the Gofar transform fault, \citeA{froment_imaging_2014} observed, after an earthquake, a drop in seismic velocities followed by a partial recovery and attributed it to coseismic-induced damage and subsequent healing mechanisms. \textcolor{black}{A similar process has been observed by \citeA{brenguier_postseismic_2008} at Parkfield with a healing phase of approximately three years.} This healing may entail microcrack closure due to stress alterations or fracture sealing driven by chemical processes \cite{mitchell_experimental_2008, brantut_time-dependent_2013}. Finally, a decrease in the damage zone width with depth has been observed both in field study \cite{cochran_seismic_2009} and inferred from numerical simulations \cite{okubo_dynamics_2019}.

Numerous studies have recently highlighted the strong thermo-hydro-mechano-chemical (THMC) coupling that exists between faults and their surrounding environments. For example, geometrical complexity exerts a strong influence on the seismicity \textcolor{black}{\cite{barnes_slow_2020, xu_fault_2023, morad_fault_2022, bedford_fault_2022, palgunadi_rupture_2024, mia_rupture_2024}}. 
The numerical study by \citeA{romanet_fast_2018} demonstrated that slow slip events (SSEs) and regular earthquakes emerge at the same location from a simple two overlapping faults system with spatially constant rate-and-state parameters, showing that fault geometrical complexity induces complex slip dynamics. 
In their analysis of 27 earthquakes, \citeA{perrin_location_2016} concluded that the degree of fault damage correlates with the amount of slip during an earthquake. 
When a fault slips the off-fault damage induces a change in bulk physical properties which in turn significantly affects the fault behavior. Damage-induced changes in elastic properties not only alter rupture extension and dynamics \cite <e.g.,>[and others]{thomas_effect_2017}, but also influence permeability, leading to changes in pore pressure that subsequently affect the fault's resistance to slip. In their study of the Gofar transform fault, \citeA{froment_imaging_2014} demonstrated a covariation between seismicity rates and changes in seismic velocity over time due to healing processes. Additionally, the evolution of the bulk medium plays a crucial role in the stored energy within it. The damage zone, for instance, can serve as an energy sink for off-fault energy dissipation \textcolor{black}{\cite{andrews_rupture_2005, johnson_energy_2021}}. Furthermore, contrary to previous beliefs, recent modeling by \citeA{okubo_modeling_2020} revealed that the damage zone significantly contributes to the overall energy budget, even at depths where its width narrows but crack density increases. Finally, time evolution of bulk properties can affect the mode of slip, as we further demonstrate in this paper. 

\par Although numerous geological, geodetical and numerical studies have highlighted the coupling between the bulk and the fault plane, they are still studied as independent entities. Prevailing modeling strategies typically assume that the effects of the damage zone can be disregarded, particularly as it narrows with depth. However, the objective of this study is to challenge this assumption by examining the significance of the damage zone and its impact on rupture dynamics, even in regions where it is narrower. In the initial section, we introduce the damage model characteristics and parametrization. In the second section, we explore the evolution of damage zone characteristics with increasing depth. The third section is dedicated to examining the influence of the damage zone on the dynamics of rupture. Finally, we introduce in the discussion a simplified "proof of concept" model that integrates a secondary cycle along the traditional seismic cycle, to discuss how the dynamic evolution of fault zone properties with slip behavior can impact the deformation modes. This approach highlights the importance of taking into account the fault zone structure inheritance into the seismic cycle. It emphasizes the necessity to account for the evolving properties of both the fault core and the surrounding bulk material, underscoring their complex interplay and significance for seismic hazard assessment and mitigation strategies.


\section{Materials and Methods}
\label{sec:mat_meth}

\par The damage zone as a function of depth is studied using a constitutive damage model implemented in the 2D fully dynamic spectral element code SEM2DPACK \cite{ampuero_sem2dpack_2012}.
The micromechanical model used in this study accounts for dynamic evolution of elastic moduli at high-strain rates and includes a physical crack-growth law to model the evolution of damage. This enables the modeling of the feedback between off-fault damage and seismic rupture. The model uses an energy-based approach to determine the nonlinear constitutive strain-stress relationship of a damaged solid. In other words, fracture damage is accounted for by creating an energetically equivalent solid.
Below only key characteristics of the model are presented, but a more detailed description can be found in \citeA{bhat_micromechanics_2012} or \citeA{thomas_dynamic_2018}.

\subsection{Constitutive Laws for Damage Modeling}
\label{sec:constitutive_laws}

\par
\begin{figure}[h!]
    \centering
    \includegraphics[width=\textwidth]{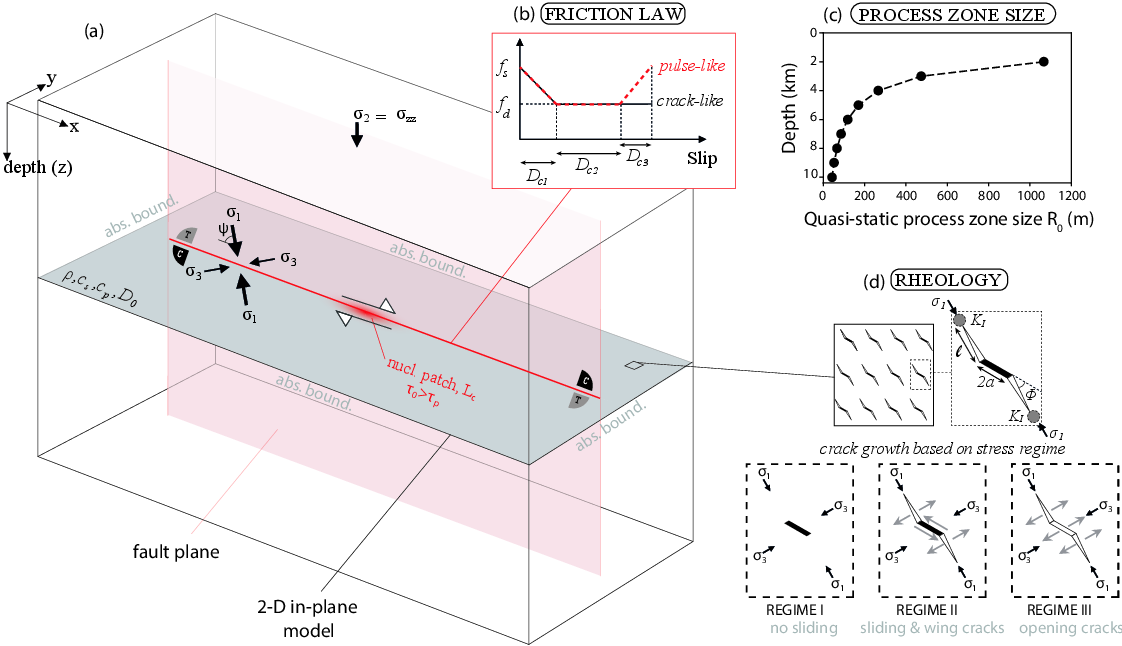}
    \caption{\textbf{(a)} Schematic 3D model: \textcolor{black}{1D} projection (red) of a 2D strike-slip fault (salmon) onto a 2D in-plane model (gray). The fault is governed by a slip weakening restrengthening friction law. $T$ and $C$ denotes tensional and compressional quadrants respectively. \textcolor{black}{\textbf{(b)} Slip-weakening-restrengthening law used (red).} \textbf{(\textcolor{black}{c})} Quasi-static process zone size $R_0$ as a function of depth. \textbf{(\textcolor{black}{d})} Regimes for inelastic deformation.}
    \label{fig:schematic}
    \end{figure}

 \textcolor{black}{\cref{fig:schematic} presents the model overview.} The medium surrounding fault is represented as an isotropic elastic solid containing pre-existing monosized flaws in the form of penny-shaped cracks of radius $a$ with a volume density $N_v$ remaining constant which implies no nucleation of new cracks. Only cracks optimally oriented from a Coulomb friction perspective are considered, meaning cracks aligned to $\sigma_1$ at the angle $\Phi = \frac{1}{2} \tan^{-1}(1/f_c)$, where $f_c$ is the friction coefficient and $\sigma_1$ is the largest stress component. The initial damage state $D_0$, representing the density of initial flaws per unit volume, is given by:
\begin{equation}
    D_0 = \frac{4 \pi}{3} N_v (a \cos \Phi)^3,
\end{equation}
with $a \cos \Phi$ the projection of the crack radius on $\sigma_1$.
\par In the model, inelastic deformation occurs through the opening and/or propagation of pre-existing cracks. These cracks grow parallel to $\sigma_1$ in the form of tensile wing-cracks of length $l$ that nucleate at the tips of the penny-shaped flaws. Consequently, the current damage state, representing the fraction of volume occupied by micro-cracks and reflecting the inelastic state of the solid is:

\begin{equation}
    D = \frac{4 \pi}{3} N_v (a \cos \Phi + l)^3.
    \label{eq:damage_state}
\end{equation}
Here, $D \in [D_0, 1]$, with $D = 1$ indicating the coalescence stage that leads to the macroscopic failure of the solid. The damage state increases as cracks grow following a state evolution law derived by differentiating \cref{eq:damage_state} with respect to time:

\begin{equation}
    \frac{dD}{dt} = \left(\frac{3 D^{2/3} D_0^{1/3}}{a \cos \Phi}\right) \frac{dl}{dt},
    \label{eq:evolution_law}
\end{equation}
with $dl/dt$ the instantaneous wing-crack tip speed.

\par At each time step, depending on the local state of stress within a cell of the simulated domain, three different regimes can be reached \textcolor{black}{(\cref{fig:schematic}d)}. Under compressive loading Regime I, stresses are not sufficiently high to induce sliding along microcracks and the solid behaves like an isotropic linear elastic material. Still for compressive loading, Regime II is reached when the shear stress $\tau$ overcomes the frictional resistance $f_c(-\sigma)$ acting on microcracks. In this regime, inelastic deformation is accounted for by the growth of tensile wing cracks at the tip of the penny-shaped cracks. Under tensile loading Regime III, both penny-shaped cracks and wing-cracks can open. 
\par Crack growth depends on both the local stress conditions at the crack tip and the material's ability to resist fracture. In the model, a crack grows if the dynamic microcrack stress intensity factor $K_I^d$ overcomes the material resistance to fracturing, given by the material dynamic initiation toughness $K_{IC}^D$ $(K_I^d \geq K_{IC}^D)$. Once the crack is initiated, the crack growth is controlled by the dynamic propagation toughness $K_{IC}^d$. Then we obtain from the above-mentioned a nonlinear equation for the crack-tip speed $dl/dt$ that can be used in \cref{eq:evolution_law} to solve for damage evolution. Experiments have shown that fracture toughness of rock is rate dependent. Therefore, under high loading rates approaching coseismic conditions, it is more difficult to initiate and propagate cracks \cite{wang_measurement_2011, zhang_effect_2013, gao_investigation_2015}. This rate dependency of fracture toughness represents the key ingredient of this damage model, compare to other micromechanical models.

\par Following the formalism of equilibrium thermodynamics and leveraging the properties of thermodynamic potential, \textcolor{black}{we recomputed the homogenized constitutive relationship between stress and stain and its evolution by the state variable $D$} \cite{rice_inelastic_1971}. \textcolor{black}{A detailed description of the model is provided in the supplementary material of  \citeA{thomas_dynamic_2018}.
}


\subsection{Non-Dimensionalization}
\label{sec:non_dimensionalization}

\par The system is non-dimensionalized by the quasi-static process zone size $R_0$ which characterizes the length scale of the frictional weakening process (and hence provides a guideline for numerical resolution) \cite{poliakov_dynamic_2002, rice_off-fault_2005}. It is given by:
\begin{equation}
    R_0(z) = \frac{9 \pi \mu G}{16 (1 - \nu) \left[(f_s - f_d) (- \sigma_{yy}^0(z)) \right]^2},
    \label{eq:R0}
\end{equation}
with $\mu$ the shear modulus, $G$ the fracture energy, $\nu$ the Poisson's ratio and $\sigma_{yy}^0$ the initial stress normal to the fault.

Therefore, the grid size depends on $R_0$ to maintain a consistent process zone resolution for simulations at different depths. As $R_0$ decreases with depth (\cref{fig:schematic}c) as a function of $\sigma_{yy}^0(z)^{-2}$, the grid size also decreases. In all simulations the process zone is resolved using 20 nodes. Fig.~S1 highlights the importance of non-dimensionalization.

\par The time step $\Delta t$ is chosen according to a Courant–Friedrichs–Levy (CFL) condition: $\Delta t = C \Delta x / c_p$, where $C$ is the CFL constant, $\Delta x$ the grid size and $c_p$ the longitudinal wave speed. Between simulations at various depths, the sole parameter modified is the grid size $\Delta x$ which linearly depends on $R_0$. Therefore, the time step is inherently dependent on the parameter $R_0$.
    
\subsection{Numerical Method and Model Set Up}
\label{sec:numerical_method}

\par We consider a 2D in-plane model \textcolor{black}{(gray plane in \cref{fig:schematic}a)} with a 1D \textcolor{black}{infinite} right-lateral fault \textcolor{black}{(red line)} at a prescribed depth. Rupture is promoted using a nucleation prone patch in the middle of the fault where the initial shear stress is slightly above the fault strength (0.01\%). The length of this patch is taken to be 10\% above the minimum nucleation length $L_{nuc}$ defines as \cite{palmer_growth_1973}:

\begin{equation}
    L_{nuc} = \frac{64}{9 \pi^2}(1 + S)² R_0,
\end{equation}
where the parameter $S = (\tau_p - \tau_0) / (\tau_0 - \tau_r)$, as defined by \citeA{andrews_rupture_1976}, gives the threshold at which seismic rupture becomes supershear ($S < 1.77$). The terms $\tau_p = -f_s \textcolor{black}{\sigma_{yy}}$ and $\tau_r = - f_d \textcolor{black}{\sigma_{yy}}$ are the peak and residual strength, respectively, and $\tau_0$ is the initial shear stress. 

\par We use a slip-weakening law with restrengthening (\cref{fig:schematic}b) to generate a pulse-like rupture. As a consequence, over a certain distance, the fault strength increases up to its static value preventing the fault to slip further. Thus, even as the rupture propagates, only a small portion of the fault is sliding at a given time. This allows for a fair comparison between simulations by avoiding the effect of having different rupture duration.
This distance is directly proportional to $D_c$ which varies with depth, or more precisely with $\sigma_{yy}$ (see \citeA{jeandet_ribes_importance_2023} for its relation to depth) as:

\begin{equation}
    D_c = \frac{2 G}{\left(f_s - f_d\right)(-\sigma_{yy}(z))}.
	\label{eq:G}
\end{equation}
where $G$, the fracture energy required to break the contact connection \cite{palmer_growth_1973}, is assumed to remain constant with depth. 
\par Uniform background stresses are applied with the maximum compressive stress $\sigma_1$ forming a 60$^\circ$ angle with the fault plane. The initial normal and shear stresses are uniform along the fault, except for the nucleation patch. 2D simulations are performed under the plane-strain assumption but the initial stress field is set up in 3D to ensure a correct stress field as emphasized by \citeA{jeandet_ribes_importance_2023}. Under the plane-strain assumption, the initial stress state is given by:
\begin{equation}
    \sigma_{ij}^0 = 
    \begin{pmatrix}
    \sigma_{xx}^0 & \sigma_{xy}^0 & 0\\
    \sigma_{xy}^0 & \sigma_{yy}^0 & 0\\
    0 & 0 & \sigma_{zz}^0
    \end{pmatrix}.
\end{equation}

Assuming a parameter $S$ and the friction coefficients $f_s$ and $f_d$, one can get the initial friction $f_0$:
\begin{equation}
    f_0 = \frac{f_s + S f_d}{1 + S} = \frac{\sigma_{xy}^0}{-\sigma_{yy}^0}. 
\end{equation}

The vertical stress $\sigma_{zz}^0$ depends linearly on depth following:
\begin{equation}
    \sigma_{zz}^0 = \rho g z (1 - \lambda),
\end{equation}
with $\rho$ the density of rock, $g$ the gravitational acceleration, $z$ the depth measured from the surface and $\lambda$ the pore pressure ratio. \textcolor{black}{We assume a constant $\lambda$ with depth although this may not hold true due to a potential transition from hydrostatic to lithostatic conditions.  Such a transition could influence the damage zone width and requires more careful investigation. However, in this model, we aim to minimize the number of parameters.} Additionally, $\sigma_{zz} = \nu (\sigma_{xx} + \sigma_{yy})$. Assuming an angle $\Psi$ between the most compressive stress $\sigma_1$ and the fault plane, the following equality holds:

\begin{equation}
	\frac{2 f_0}{\tan \left(2 \Psi\right)} + 1 = \frac{\sigma_{xx}^0}{\sigma_{yy}^0}.
\end{equation}

This allows to derive the last stress matrix components:

\begin{subequations}
	\begin{equation}
		\sigma_{yy}^0 = \frac{\sigma_{zz}^0}{\nu \left(\frac{2 f_0}{\tan \left(2 \Psi\right)} + 2\right)},
	\end{equation}
	\begin{equation}
		\sigma_{xx}^0 = \left(\frac{2 f_0}{\tan \left(2 \Psi\right)} + 1 \right) \sigma_{yy}^0,
	\end{equation}
	\begin{equation}
		\sigma_{xy}^0 = \left(\frac{f_s + S f_d}{1 + S}\right) \sigma_{yy}^0.
	\end{equation}
\end{subequations}

\par The domain size is set to be large enough so that waves do not have time to bounce back from the boundaries and interact with the propagating rupture.

\subsection{Closeness to Failure}

\par For a fair comparison, the closeness to failure for off-fault cracks must be equal at all depths. In our model, this equates to defining our proximity to reaching Regime II. This implies that the initial stress intensity factor of the bulk must remain constant across all simulations. The determination of this consistency hinges on the initial stress state\textcolor{black}{, which is strictly governed by depth as shown in \cref{sec:non_dimensionalization}}, and on the friction coefficient on the microcracks, $f_c$. Consequently, $f_c$ varies with depth while adhering to experimental values ($f_c \in [0.7, 0.84]$, see \cref{tab:mus_param}). It is important to note that we do not assert that the static friction coefficient naturally varies with depth. Instead, our objective is to facilitate a fair comparison by isolating the influence of depth on the generation of the damage zone.

\begin{table}[h!]
    \centering
    \begin{tabular}{|c|c|}
		\hline
		\textbf{Depth (km)} & \textbf{Bulk static friction coefficient} \\ 
  		\hline
         2 & 0.84 \\
         3 & 0.78 \\
         4 & 0.75 \\
         5 & 0.73 \\
         6 & 0.72 \\
         7 & 0.71 \\
         8 & 0.71 \\
         9 & 0.70 \\
         10 & 0.70 \\
		\hline
    \end{tabular}
    \caption{Bulk static friction coefficient at each studied depth.}
    \label{tab:mus_param}
\end{table}

\par All parameters values used are given in \cref{table:param}.

\begin{table}[h!]
	\centering
	\begin{tabular}{ |c|c|c| } 
		\hline
		\textbf{Symbol} & \textbf{Parameter} & \textbf{Value} \\ 
		\hline 
		$\mu$ & Shear modulus & 26.2 GPa  \\ 
		$\nu$ & Poisson's ratio & 0.276 \\ 
		$\rho$ & Density of rock & 2700 $\mathrm{kg/m^3}$ \\
		 \textcolor{black}{$\Psi$} & Orientation of $\sigma_1$ & 60$^\circ$ \\
		$S$ & Seismic ratio & 1 \\
		$\lambda$ & Lithostatic pore pressure & 0.4 \\
		$f_c$ & Bulk friction coefficient & [0.7, 0.84] \\
  		$f_s$ & Fault static friction coefficient & 0.6 \\
		$f_d$ & Fault dynamic friction coefficient & 0.1 \\
		$D_c$ & Critical slip distance & from \cref{eq:G} \\
        $D_{c2}$ & Critical slip distance 2 & $5 \times D_c$ \\
        $D_{c3}$ & Critical slip distance 3 & $0.1 \times D_c$ \\
		$G$ & Fracture energy & $\approx$ 22 $\mathrm{MJ/m^2}$ \\
		$D_0$ & Initial damage state & 0.1 \\
		$c_s$ & S-wave speed & 3115 m/s \\
		$c_p$ & P-wave speed & 5600 m/s \\		
		\hline
	\end{tabular}
	\caption{Parameters used for simulations with SEM2DPACK.}
	\label{table:param}
\end{table}

\section{Evolution of Damage Zone Width and Density With Depth} 

\par
\begin{figure}[h!]
    \centering
    \includegraphics[width=\textwidth]{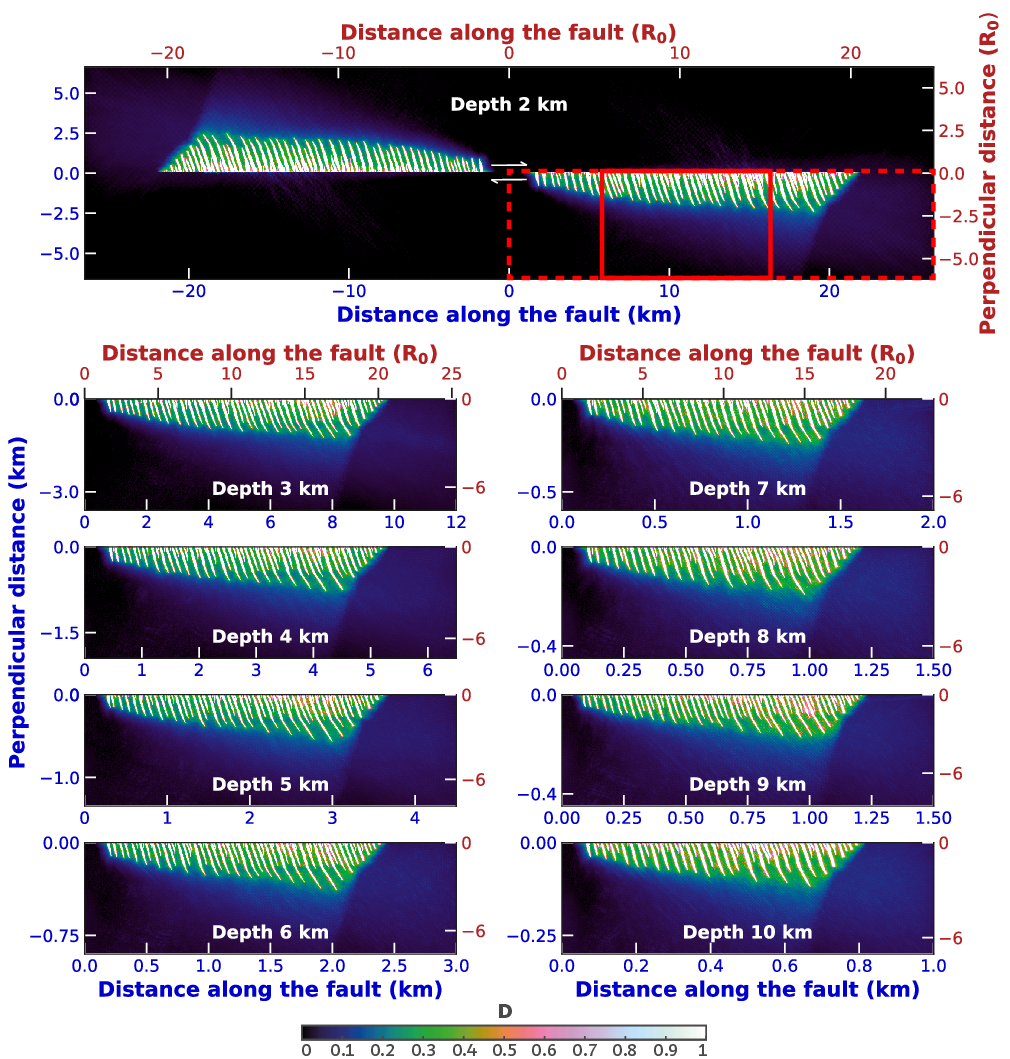}
    \caption{Comparison of damage arising from a rupture at depths ranging from 2 to 10~km. The top panel depicts the entire fault at a depth of 2~km. The small red rectangle demarcates the region from which densities in \cref{fig:width_and_profile} are computed. Given the symmetry of the rupture, the subsequent panels exclusively showcase the bottom-right quadrants (as delimited by the dashed red rectangle in the top panel) of the domain. Both real (km) and non-dimensionalized ($R_0$) distances are given.}
    \label{fig:all_depths}
\end{figure}

 To explore the evolution of damage zone width and density with depth, simulations were conducted at 1~km depth intervals ranging from 2 to 10~km. \cref{fig:all_depths} depicts the damage state at the end of each simulation for the respective depths. Given the symmetry of the rupture with respect to the fault center, only one quarter of the domain is presented except for the simulation at 2~km depth. 

\par
\begin{figure}[h!]
    \centering
    \includegraphics[width=\linewidth]{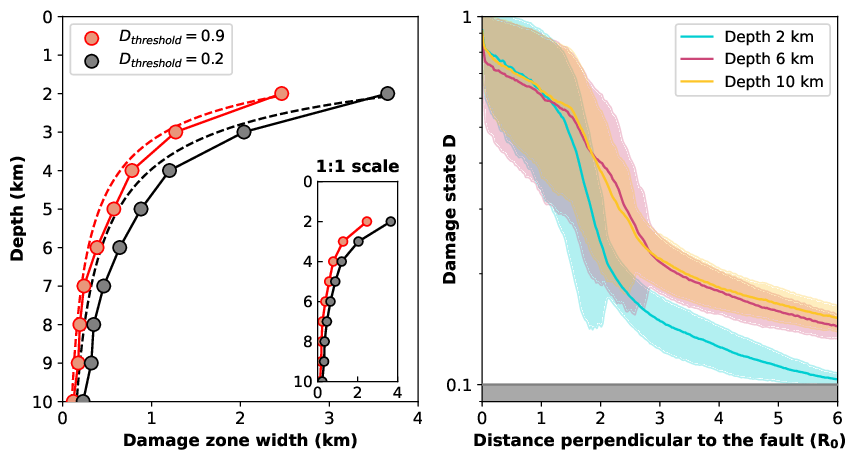}
    \caption{\textbf{Left:} Damage zone width as a function of depth for two distinct damage thresholds, revealing a consistent funnel-shaped structure. The dashed lines indicate a fit by $c \ R_0$, with $c$ a positive constant ($\approx$ 2.3 and 3.4 for the threshold 0.9 and 0.2 respectively). The inset displays the same scale for both width and depth. \textbf{Right:} Comparison of damage density with perpendicular distance from the fault for simulations at 2, 6 and 10~km depth. Computations are done in a region between 10 and 15 $R_0$ away from the fault center to avoid nucleation effects. The location is shown in \cref{fig:all_depths} as a red rectangle. Bright colors represent the average damage state for a given distance perpendicular to the fault, while shaded areas indicate the standard deviation. The grey area marks the initial damage state $D=0.1$.  Note the logarithmic scale for the damage state.}
    \label{fig:width_and_profile}
\end{figure}

 At a first glance, the damage states at different depth look similar: they all develop highly damaged elongated branches with $D$ close to 1 and a maximum length of approximately $2.5 \ R_0$\textcolor{black}{, making an angle of $\approx 60-70$° which is close to $\Psi$}. However, as $R_0$ decreases drastically with depth (\cref{fig:schematic}c), there is a corresponding diminution in the extent of the damage zone. The left panel of \cref{fig:width_and_profile} shows the evolution of the maximum damage zone width with depth, defined as the furthest distance from the fault at which $D \geq D_{thres}$. Notably, regardless of the chosen $D_{thres}$, the evolution of the damage zone width with depth reveals a funnel-shaped structure. Note that to prevent potential confusion with a term employed in tectonics to denote a particular architecture of strike-slip systems, we have opted not to employ the term “flower structure” despite its occasional usage in the literature. We instead humbly suggest ``funnel-shaped structure''. This structure aligns well with geophysical observations of \citeA{cochran_seismic_2009} and is consistent with the findings from simulations conducted by \citeA{okubo_dynamics_2019}. Their approach to model damage differs from ours: coseismic off-fault damage is accounted for with nucleation of new cracks but elastic properties are kept constant. In our approach there is no nucleation of new cracks, but damage is accounted for by computing the change in elastic properties due to crack growth. It is satisfying to see that both methods lead to qualitatively similar conclusions. 

\par In the right panel of \cref{fig:width_and_profile}, we compare the average damage states with perpendicular distance from the fault at depths of 2, 6 and 10~km. Bright colors represent the average damage state, while shaded areas indicate standard deviation. It is clear that, away from the fault, at distances higher than $\approx 2.5 \ R_0$, a distance that aligns with the size of the branches, the damage density increases with depth. Regions closer to the fault exhibit a more ambiguous trend; however, within this region, the value of the standard deviation increases with depth, indicating a higher overall damage state. The diminished distinction closer to the fault can be attributed to the saturation of the damage state $D$ to 1, a limitation inherent to our model.

\par
\begin{figure}[h!]
    \centering
    \includegraphics[width=\linewidth]{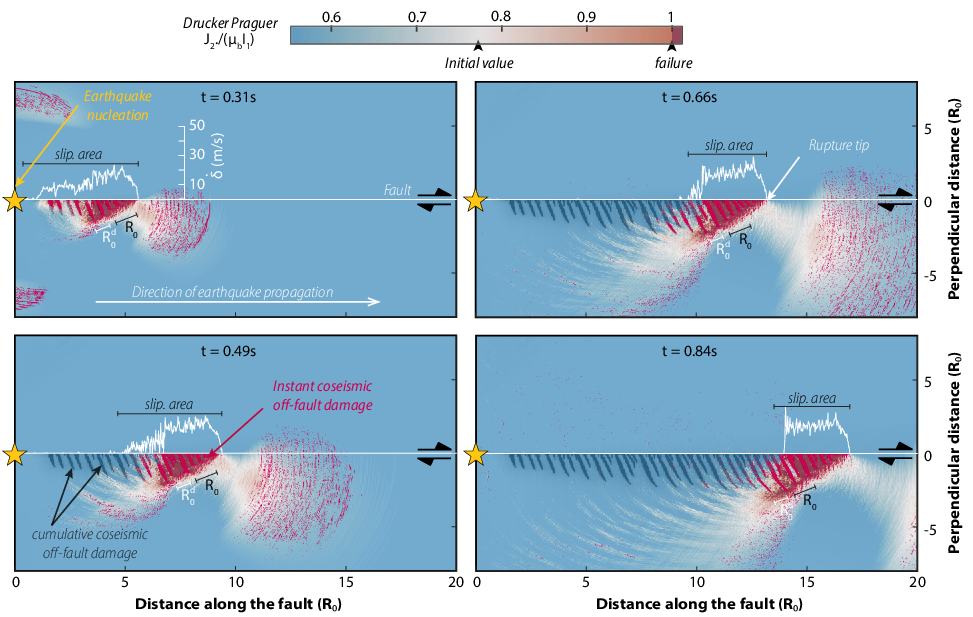}
    \caption{Temporal evolution of a dynamic rupture at 6~km depth. Colors indicate the normalized Drucker-Prager yield criterion as reference. Fault slip rates (white curves) highlight the slipping area, and cumulative damage $D$ (grey scale) is superimposed. Instantaneous coseismic off-fault damage is underlined in red. The static and dynamic process zone sizes \textcolor{black}{$R_0$} and $R_0^d$ are indicated. \textcolor{black}{$R_0^d$ is numerically calculated as the distance behind the crack tip where the slip is equal to $D_c(z)$.} The yellow star denotes the nucleation patch.} 
    \label{fig:DP}
\end{figure}

 \cref{fig:DP} illustrates the rupture propagation at a specific reference depth of 6 km. The Drucker-Prager yield criterion in the background delineates areas where coseismic off-fault damage predominantly occurs. Additionally, the graph displays the fault slip rate (in white) at the time of the snapshot, cumulative damage (in grey) accumulated so far, and off-fault damage during this particular time step (in red). Coseismic off-fault damage primarily occurs behind the rupture front, where the fault slip. The damage occurring behind the area which is currently slipping is a result of cascading effects associated with the growth of these branches, altering the stress state ahead of them. Propagating waves in the medium are also playing a role. Notably, the spacing between branches \textcolor{black}{seems to correspond} primarily to the dynamic process zone size, reflecting the intricacies of the dynamic rupture process.

\medskip

\par The elongated branches observed herein manifest as a consequence of a single rupture within a medium characterized by a homogeneous initial damage state. 
This contrasts with field observations, where the signature of multiple ruptures persists. They show an exponential \cite{vermilye_process_1998, wilson_microfracture_2003, faulkner_slip_2006, faulkner_scaling_2011, mitchell_nature_2009} or power law \cite{savage_collateral_2011, rodriguez_padilla_accrual_2022} decay in the damage state with increasing distance from the fault. Consequently, we chose to run additional simulations with an initial damage state decreasing exponentially from $D_0 = 0.6$ to $D_0 = 0.1$ over a distance $3 \ R_0$. Fig.~S2 shows the corresponding damage state at the end of each simulation for the respective depths. The damage zones exhibit comprehensive regions characterized by heightened damage, as opposed to isolated branches in the case of an homogeneous initial damage state. Even in this case, the evolution of the maximum damage zone width with depth has a funnel-shaped structure as shown in Fig.~S3. 

\par For both an homogeneous initial damage stage and an exponentially decreasing one, the damage zone narrows with depth in correlation with the reduction of the process zone size. Despite the relatively small width of the damage zone, its influence on rupture dynamics remains significant, given that all processes occur at the scale of the process zone. The next section aims to illustrate the damage zone influence on rupture dynamics. 

\section{Damage Influence on Rupture Dynamics}

\begin{figure}[h!]
    \centering
    \includegraphics[width=\textwidth]{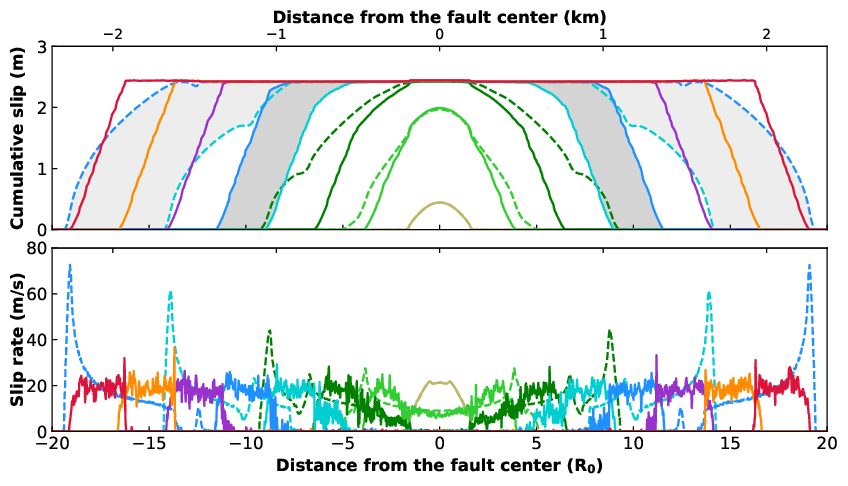}
    \caption{Comparison of cumulative slip (top) and slip rate (bottom) as a function of distance from the fault center for both the elastic (dashed lines) and damage (continuous lines) cases at 6~km depth. Colors are isochrones, the dark and light gray areas highlight the area that slipped during a given time interval for the damage and elastic case respectively. Both real (km) and non-dimensionalized ($R_0$) distances are given.}
    \label{fig:cum_slip_and_slip_rate}
\end{figure} 

\par \cref{fig:cum_slip_and_slip_rate} compares the cumulative slip and the slip rate for the elastic and damage cases at 6~km depth. In the elastic scenario, the slip rate keeps growing as slip accumulate, as predicted theoretically for the law we are using. However, in the presence of damage, the slip rate remains approximately constant, effectively mitigating the emergence of nonphysical singularities in the slip rate. The oscillations observed in the slip rate for the damage case are attributed to the generation of high-frequency waves \cite{thomas_effect_2017}. The cumulative slip illustrates in both cases the pulse-like nature of the rupture consistent with the employed slip-weakening-restrengthening law. The shaded dark and light grey regions highlight the slipping portion of the fault during the same time interval for the damage and elastic cases respectively, showing a broader pulse width for the elastic case. This is related to the fact that the ruptures become supershear for the purely elastic cases (for a more detailed discussion, please refer below). These findings persist across all depths as shown in Fig.~S4 and Fig.~S5. For both the damage and elastic cases, the slip rate increases with depth.

\begin{figure}[h!]
    \centering
    \includegraphics[width=\textwidth]{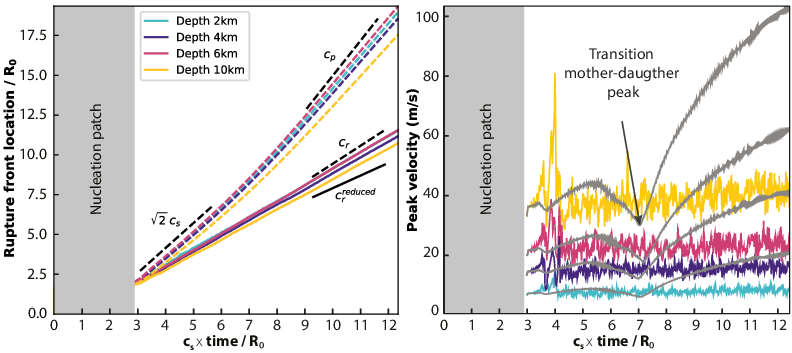}
    \caption{\textbf{Left:} Comparison of rupture front locations as a function of dimensionless time at different depths both for elastic (dashed lines) and damage (continuous lines) cases. The black dashed lines indicate for reference the speed domain limits for stable supershear rupture, $\sqrt{2}c_s$ and $c_p$, as well as the upper limit for subshear rupture $c_r$, where $c_s$, $c_p$ and $c_r$ are respectively the shear, longitudinal and Rayleigh wave speeds. The continuous black line illustrates the reduced Rayleigh wave speed (decreased by 30\%) due to damage. \textbf{Right:} Comparison of peak velocity in $m/s$ as a function of dimensionless time at different depths both for elastic (dashed grey lines) and damage (continuous lines) cases.}
    \label{fig:rupture_front_peak_velocity}
\end{figure} 

\par \cref{fig:rupture_front_peak_velocity}a compares the evolution of rupture front locations over time at 2, 4, 6 and 10~km depth for both elastic and damage cases. In the damage scenarios, rupture speeds remain approximately constant and are consistently lower than those observed in the corresponding elastic cases: damage generation slows the rupture speed at all depths. In the elastic case, the rupture swiftly transitions to supershear as highlighted, while in the damage case, it consistently remains within the subshear domain. Damage generation thus delays, if not entirely prevents, the transition to supershear at all depths in these cases. Therefore, the existence of a damage zone could explain why supershear ruptures are not as ubiquitous in nature. It is worth pointing that \citeA{thomas_dynamic_2018} observed a supershear transition in the presence of damage using the same model but under different settings on a smaller fault. Understanding the role of damage in the supershear transition will therefore necessitate further studies. It is noteworthy that in our case this transition to supershear occurs at the same non-dimensionalized time across all depths, suggesting a dependence on the process zone size.

\par
\begin{figure}[h!]
    \centering
    \includegraphics[width=\textwidth]{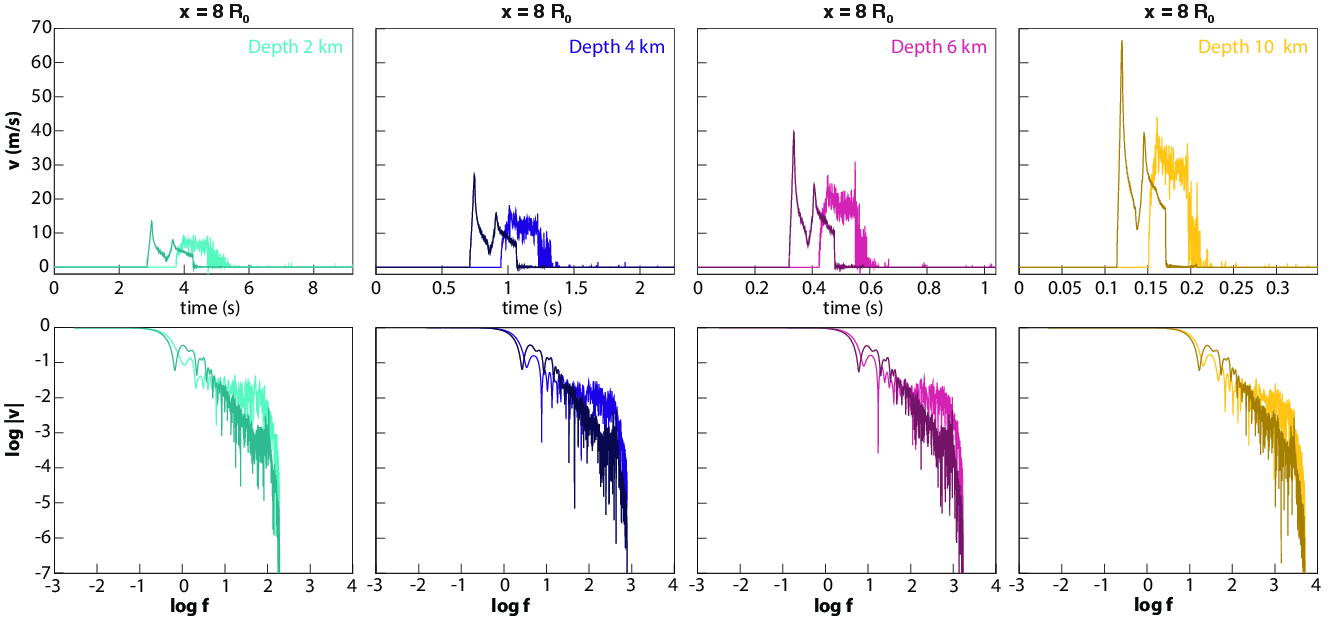}
    \caption{Comparison of Fourier amplitude spectra (FAS) of slip rate at different depths. The dark and light colors correspond to the elastic and damage case respectively.}
    \label{fig:freq_analysis}
\end{figure}

 The peak velocities over time (\cref{fig:rupture_front_peak_velocity}b) increase with depth for both elastic and damage scenarios. For the damage cases, these peaks remains relatively constant over time at a given depth. The oscillations in the peak velocities are attributed to the off-fault related high-frequency content (see \cref{fig:freq_analysis} for a discussion). With an initial damage state decreasing exponentially, the damage preceding the rupture front exceeds that observed in the initial homogeneous damage case, resulting in a further deceleration of the rupture front (Fig.~S6a). Conversely, in comparison to the homogeneous case, we note an escalation in peak velocity (Fig.~S6b) along with a reduction in oscillations. This observation is linked to the nature of the newly formed damage zone, which exhibits greater homogeneity.

\medskip

\par \cref{fig:freq_analysis} compares the Fourier amplitude spectra (FAS) of slip rate at various depths. Across all depths, the slip rate in the damage case exhibits a higher-frequency content than in the elastic case, as evidenced by a steeper slope in the Fourier velocity spectra beyond a specific frequency threshold. This phenomenon is primarily attributed to alterations in elastic properties caused by damage, culminating in the formation of a Low-Velocity Zone (LVZ) characterized by up to a 30\% decrease in wave speed. The resulting contrast in material properties leads to internal wave reflections that interact with the propagating rupture, producing a high-frequency content. The generated damage also radiates waves, a phenomenon observable upon closer examination of the created branches \textcolor{black}{\cite{thomas_dynamic_2018, ben-zion_seismic_2009}}. This analysis is in agreement with laboratory experiments that show that high-frequency radiation lies in part in the off-fault coseismic damage propagating behind the rupture front \cite{marty_origin_2019}\textcolor{black}{, as well as with numerical studies using pre-existing short branches \cite{ma_dynamic_2019} or damage-breakage rheology \cite{mia_dynamic_2024}.} Furthermore, the high-frequency content increases with depth, as shown by a shift of the frequency plateau toward higher frequencies and in agreement with natural observations. This trend is attributed to the \textcolor{black}{systematic decrease in branch size} with increasing depth. In cases with an initial exponential decay of damage, these features are maintained, if not more pronounced (Fig.~S7).

\section{Discussion and Conclusion}

\par We have investigated the evolution of the damage zone with depth using a numerical model that incorporates the dependence of fracture toughness on loading rate and crack-tip velocities. Our findings reveal that while the width of the damage zone decreases with depth, exhibiting a funnel-shaped structure correlated with the reduction of the process zone size, its density concurrently increases. Despite the relatively narrow width of the damage zone, its influence on rupture dynamics remains significant at all depths as all processes occur at the scale of the process zone. \textcolor{black}{Similar findings have been discussed by \citeA{mia_spatio-temporal_2022}.} In the presence of damage, the slip rate remains relatively constant, effectively preventing the emergence of nonphysical values of slip rates observed in the elastic case. The high-frequency content of the slip rate is enhanced by damage and depth. Additionally, damage slows down the rupture speed at all depths and can even prevent or delay the transition to supershear.

\par \textcolor{black}{Simulations are primarily conducted in 2D, but a more comprehensive investigation of 3D effects is necessary. These 3D effects are expected to be significant at shallow depths, with their influence diminishing substantially as depth increases. This has been partially explored by \citeA{ma_physical_2008}, where they compared 3D simulations to their 2D counterparts using Drucker-Prager plasticity. As shown in Figure 3 of their study, the 3D effects are pronounced at shallow depths but become negligible deeper.
}

\bigskip

\par To properly assess the damage zone strong influence on rupture dynamics, one should compute the energy budget to quantitatively measure the bulk energetic dissipation due to off-fault crack growth. This work is currently under development. Geophysical observations also suggest a gradual recovery of the elastic properties after an earthquake due to healing processes. Thus the bulk is constantly evolving during the co- and inter-seismic periods, and so is the quantity of energy stored and dissipated in the medium. The structure of the damage zone has also an impact on the permeability, hence on the fluid flow and consequently on the fault resistance to slip. An increase of permeability can also favor aseismic sliding by pressure-solution, which may play significant role in accommodating the afterslip recorded after large earthquakes. This can be followed by a transition back to seismic behavior due to compaction by pressure-solution \cite{den_hartog_frictional_2012}. The cumulative picture from these studies suggests that on top of the “seismic cycle” there is a superimposed “cycle” where the slip dynamics impact the bulk evolution, which in turn influences back the fault motion. \textcolor{black}{\citeA{meyer_off-fault_2024}  proposed, based on experimental analysis, that off-fault deformation can induce unstable slip by decreasing the stiffness of the surrounding rock volume. They emphasized that earthquakes are inherently volumetric processes and observed recurring cycles of localization and delocalization during a series of stick-slip events.} This intertwined dynamic should be explored in numerical model to determine its impact on the full seismic cycle. Our model only allows to simulate a single rupture and not the full seismic cycle. Therefore, as a proof of concept, we have implemented a simple 1D spring-slider model to simulate fault slip over several seismic cycles. The implementation allows for normal stress variations to reproduce changes of the load in the bulk such as seasonal hydrological loading or tides as well as bulk elastic properties variations in the form of varying spring stiffness (implementation details in SM). The adjustments in bulk elastic properties are designed to reflect earthquake-related off-fault damage and healing processes, as documented both in natural fault zones and laboratory experiments \cite{vidale_damage_2003, niu_preseismic_2008, shreedharan_preseismic_2020}.

\begin{figure}[h!]
    \centering
    \includegraphics[width=12cm]{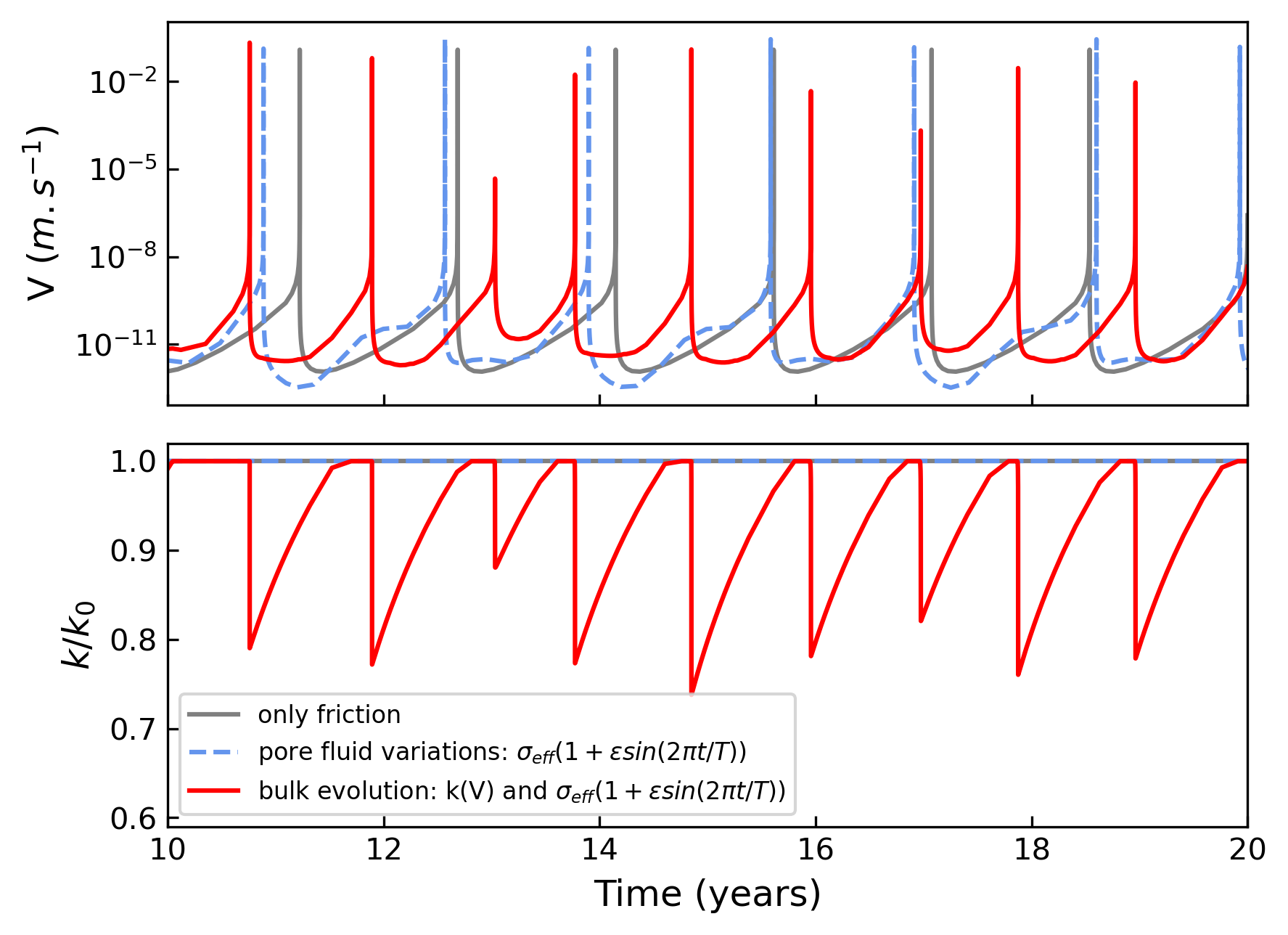}
    \caption{Comparison between cases C0, C1 and C2. C0: friction only, both normal stress and stiffness are constant (grey). C1: normal stress variation only (dashed blue). C2: normal stress and stiffness variations (red). Upper part: slip velocity as a function of time. Lower part: stiffness $k$ normalised by the initial stiffness $k_0$ as a function of time.}
    \label{fig:spring-slider}
\end{figure}

\par We compare results (\cref{fig:spring-slider}) of a reference model where the bulk properties stay constant -- both normal stress and spring stiffness constant (C0) -- with two other models: one with normal stress variations only (C1) and an other one with normal stress and stiffness variations (C2). 
In the absence of normal stress and stiffness variations (C0), the system reaches a perfectly periodic steady-state with regular earthquakes having consistent slip velocities. In model C1, sinusoidal variations in normal stress, stemming from pore pressure fluctuations, disrupt the periodicity, yet the slip velocities remain constant. These variations in pore pressure alter the effective normal stress, consequently affecting the resistance to motion and the timing of events. In C2, with both sinusoidal normal stress and stiffness variations, events become non-periodic, varying in slip velocity by several orders of magnitude. Slow slip events (SSEs) occur alongside regular earthquakes, challenging the common notion that frictional properties variations drive SSEs. Instead, in this model SSEs are attributed to a co-temporal evolution of stiffness and effective normal stress, influenced by what, in nature, would be crack healing-induced changes in bulk elastic properties,  permeability, and effective normal stress. \textcolor{black}{\citeA{mia_spectrum_2023} also observed a spectrum of slip depending on the bulk material strength, using an elastoplastic spring-slider model.}

\par This phenomena can be explained as follows. The critical nucleation size $L_c$, the minimum size for an instability to grow, is linked to a critical stiffness $k_c$ that depends on frictional parameters ($a$, $b$ and $D_c$) and effective normal stress $\sigma_{eff}$ as:

\begin{equation}
	k_c = \left|\frac{\sigma_{eff} (b-a)}{D_c}\right|.
\end{equation}
Without frictional properties variations -- as in our case -- $L_c$ solely depends on $\sigma_{eff}$ and $k$ which vary over time. Thus, within this simple 1D system, the nucleation length is sometimes small enough that an instability can grow and triggers seismic velocity while sometimes it cannot reach dynamic values. 
\par This analysis still holds for Earth's case. The stiffness $k$ can be related to elastic rock properties as follows:

\begin{equation}
	k = \frac{\mu}{(1 - \nu)L}.
\end{equation}
Then, for a slipping elliptical patch of fault length $L$ we have:

\begin{equation}
	L_c = \left|\frac{\mu D_c}{(1-\nu)\sigma_{eff}(b-a)}\right|.
	\label{eq:critical_nucleation_size}
\end{equation}
Slow slip events are explained, both in this study and in the prevailing modelling strategies, as aborted earthquake due to critical nucleation size increase. However, we differ in the suggested mechanisms behind this increase. The prevailing mechanism proposed is frictional parameters changes in which slow slip events occur on patches close to velocity neutral ($b - a \approx 0$, hence $L_c \to \infty$) while we argue that the evolution of the bulk induced by crack growth and healing, which has been abundantly observed in the field is a sufficient (and more likely?) explanation for the fault slip spectrum as observed by geodesy and seismology. \citeA{thakur_effects_2024} have conducted a comprehensive study modeling the damage zone throughout the entire seismic cycle. They modeled the damage zone as a homogeneous elastic layer surrounding the fault plane, with a lower shear modulus compared to the embedding medium (30\% reduction). Upon the initiation of coseismic slip, they applied an additional slight decrease (0.5\%) in the shear modulus, followed by a logarithmic healing process gradually restoring it to its original value over time. Their findings revealed a broad spectrum of fault slip behavior. However, they introduced friction parameter variations along the dip, unlike our 1D model, where friction remains constant. This disparity underscores that, in our scenario, it is indeed the bulk properties that dictate the response, rather than variations in friction. Although their model does not incorporate a true feedback between the bulk and the fault plane, it represents the right step towards integrating bulk evolution into seismic cycle modeling.

\section*{Open Research}
The code SEM2DPACK can be download at \url{https://github.com/jpampuero/sem2dpack}. All data are available upon reasonable request.

\section*{Acknowledgements}
This study was supported by the Agence National de la Recherche (ANR) IDEAS contract ANR-19- CE31-0004-01. The computations presented here were conducted on the MADARIAGA cluster supported by the European Research Council grant PERSISMO (grant 865411).

\clearpage

\setcounter{section}{0}
\setcounter{figure}{0}
\renewcommand\thesection{\Alph{section}}
\renewcommand\thefigure{S\arabic{figure}}
\renewcommand\thesubsection{\thesection.\arabic{subsection}}

\begin{center}
\LARGE {\bf Appendix}
\end{center}

\begin{figure}[h!]
    \centering
    \includegraphics[width=\textwidth]{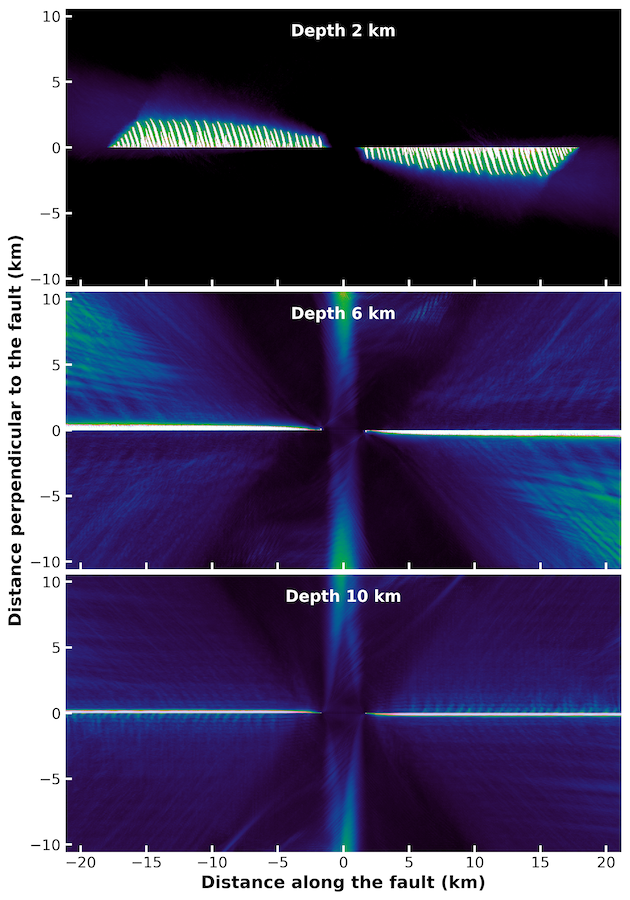}
    \caption{Comparison of damage states in simulations at 2, 6 and 10 km without non-dimensionalization. All simulations employ a consistent grid size of 53 m. Hence, there are 20 grid nodes resolving the process zone size at 2 km, while only about 2 and 1 nodes are available for simulations at 6 and 10 km, respectively. This underscores the significance of non-dimensionalization in investigating the evolution of damage states with depth.}
    \label{fig:without_adi}
\end{figure}

\begin{figure}[h!]
    \centering
    \includegraphics[width=\textwidth]{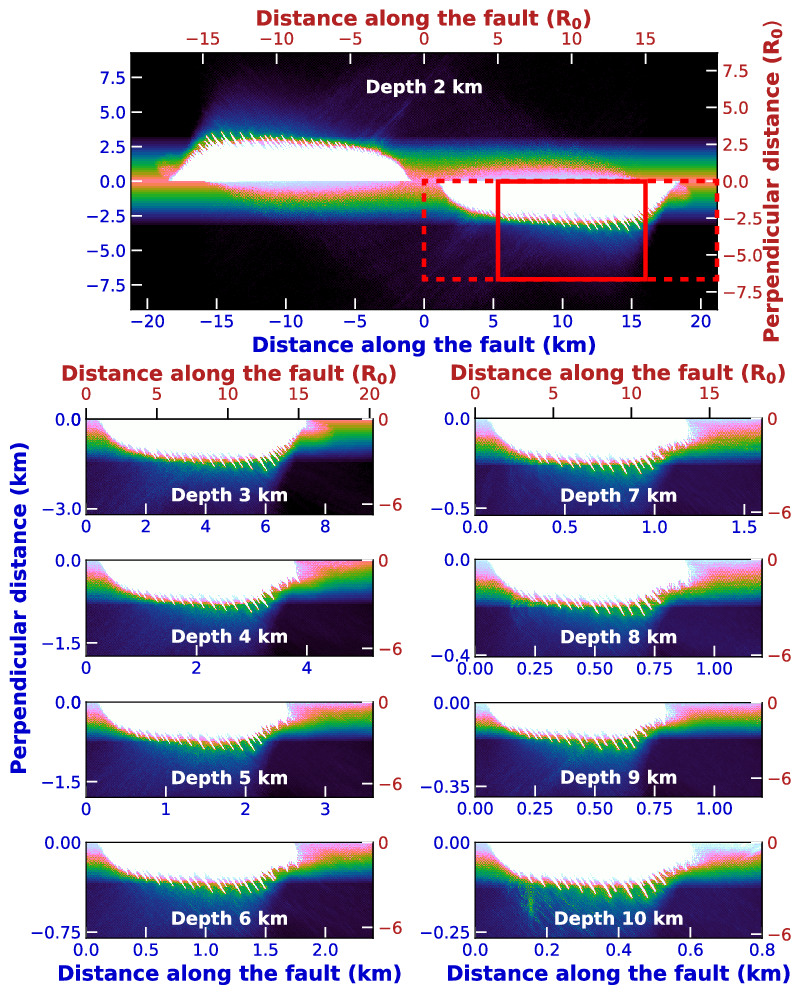}
    \caption{Comparison of damage arising from a rupture at depths ranging from 2 to 10~km with an initial damage decreasing exponentially away from the fault. The top panel depicts the entire fault at a depth of 2~km. Given the symmetry of the rupture, the subsequent panels exclusively showcase the bottom-right quadrants (as delimited by the dashed red rectangle in the top panel) of the domain. Both real (km) and non-dimensionalized ($R_0$) distances are given.}
    \label{fig:all_depths_expo}
\end{figure}

\begin{figure}[h!]
    \centering
    \includegraphics[width=\textwidth]{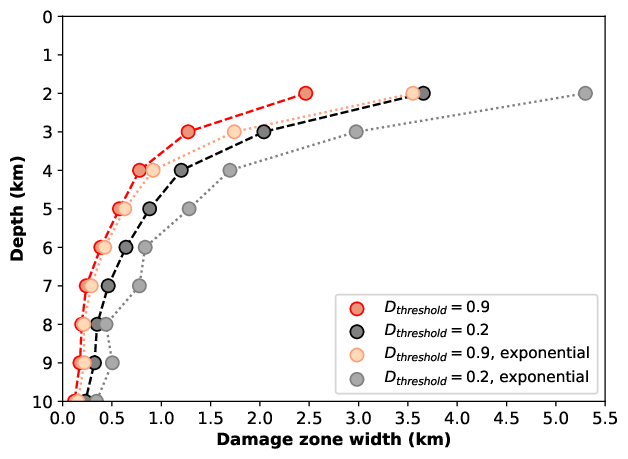}
    \caption{Comparison of damage zone width as a function of depth for two distinct damage thresholds for the cases with an initial homogeneous damage state and an initial exponential damage. Both cases display a consistent funnel-shaped structure.}
    \label{fig:width_with_expo}
\end{figure}

\begin{figure}[h!]
    \centering
    \includegraphics[width=\textwidth]{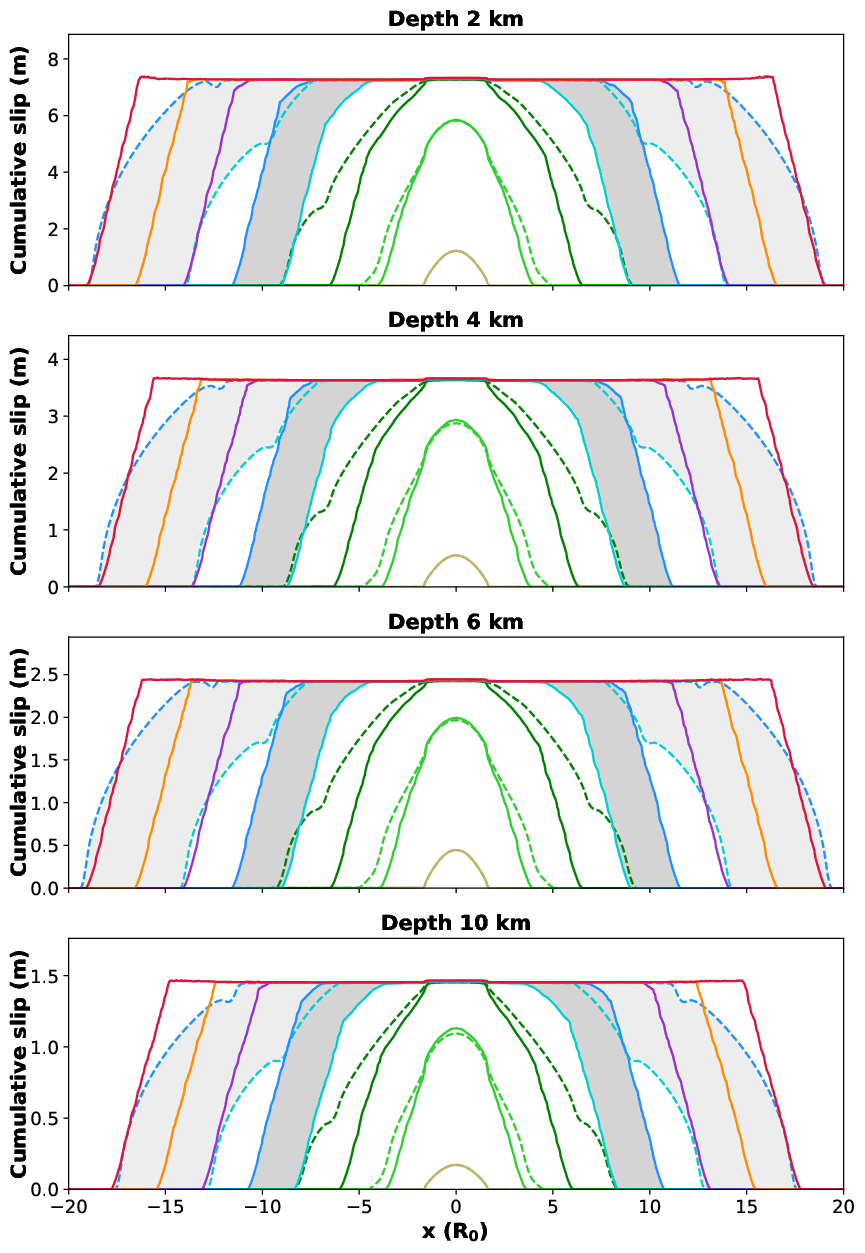}
    \caption{Comparison of cumulative slip as a function of distance from the fault center for both the elastic (dashed lines) and damage (continuous lines) cases at different depths. Colors are isochrones, the dark and light gray areas highlight the area that slipped during a given time interval for the damage and elastic case respectively. Both real (km) and non-dimensionalized ($R_0$) distances are given.}
    \label{fig:cum_slip_all_depths}
\end{figure}

\begin{figure}[h!]
    \centering
    \includegraphics[width=\textwidth]{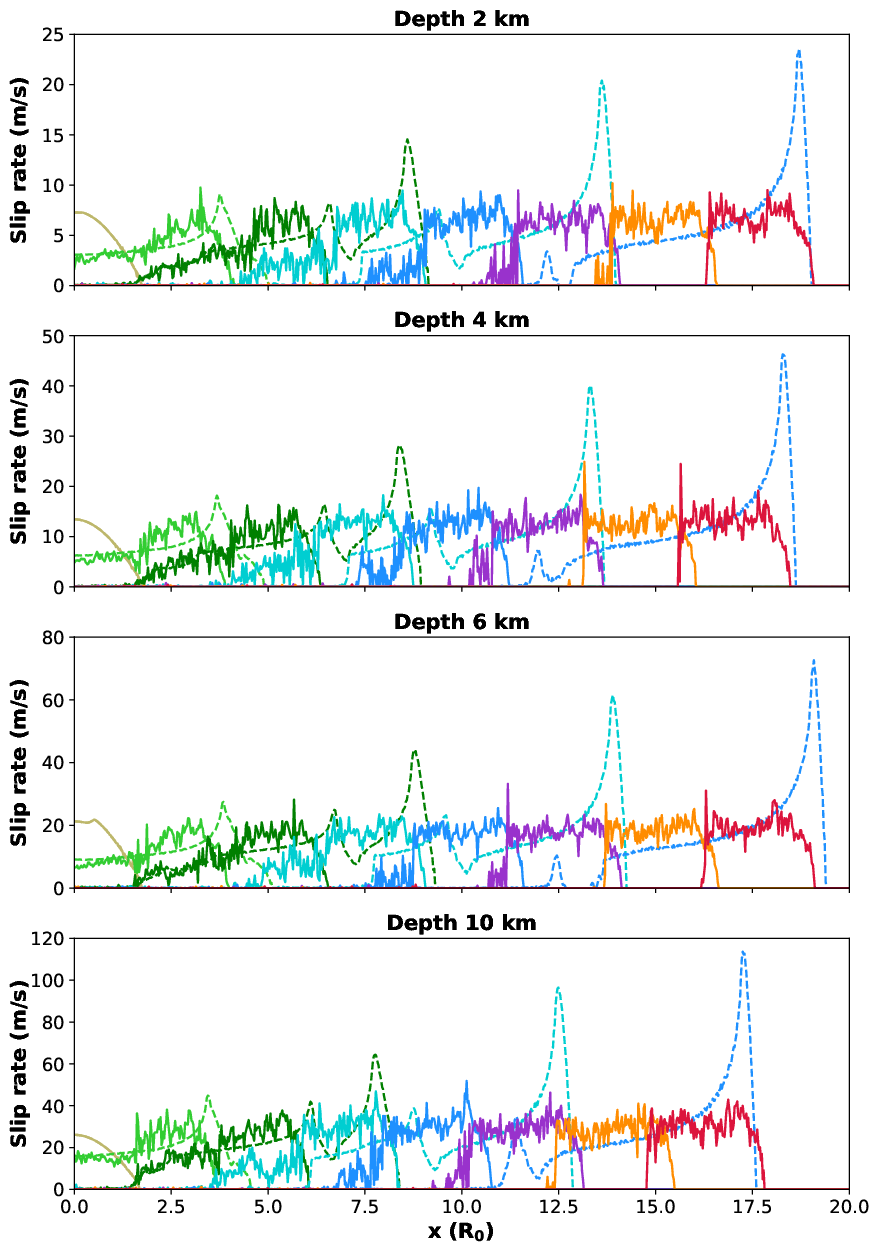}
    \caption{Comparison of slip rate as a function of distance from the fault center for both the elastic (dashed lines) and damage (continuous lines) cases at different depths. Colors are isochrones, the dark and light gray areas highlight the area that slipped during a given time interval for the damage and elastic case respectively. Both real (km) and non-dimensionalized ($R_0$) distances are given.}
    \label{fig:slip_rate_all_depths}
\end{figure}

\begin{figure}[h!]
    \centering
    \includegraphics[width=\textwidth]{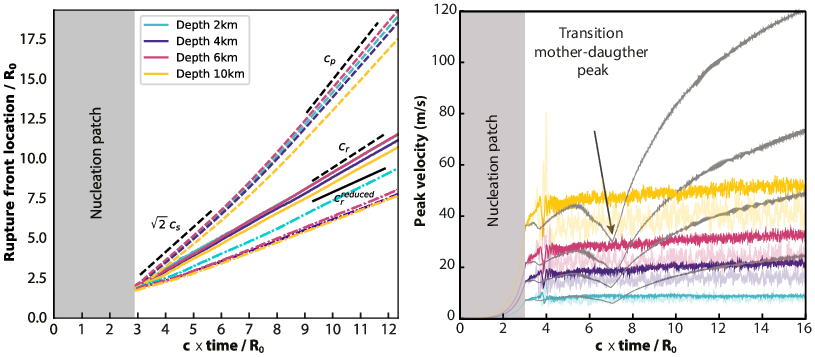}
    \caption{\textbf{Left:} Comparison of rupture front locations as a function of dimensionless time at different depths for elastic (dashed lines), homogeneous initial damage state (continuous lines) and exponentially decreasing initial damage (dashed-dotted lines) cases. The black dashed lines indicate for reference the speed domain limits for stable supershear rupture, $\sqrt{2}c_s$ and $c_p$, as well as the upper limit for subshear rupture $c_r$, where $c_s$, $c_p$ and $c_r$ are respectively the shear, longitudinal and Rayleigh wave speeds. The continuous black line illustrates the reduced Rayleigh wave speed (decreased by 30\%) due to damage. \textbf{Right:} Comparison of peak velocity in $m/s$ as a function of dimensionless time at different depths both for elastic (dashed grey lines) and exponentially decreasing initial damage (continuous lines) cases. The light continuous color lines indicate the homogeneous initial damage case for reference.}
    \label{fig:rupture_front_peak_velocity_comparison}
\end{figure}

\begin{figure}[h!]
    \centering
    \includegraphics[width=\textwidth]{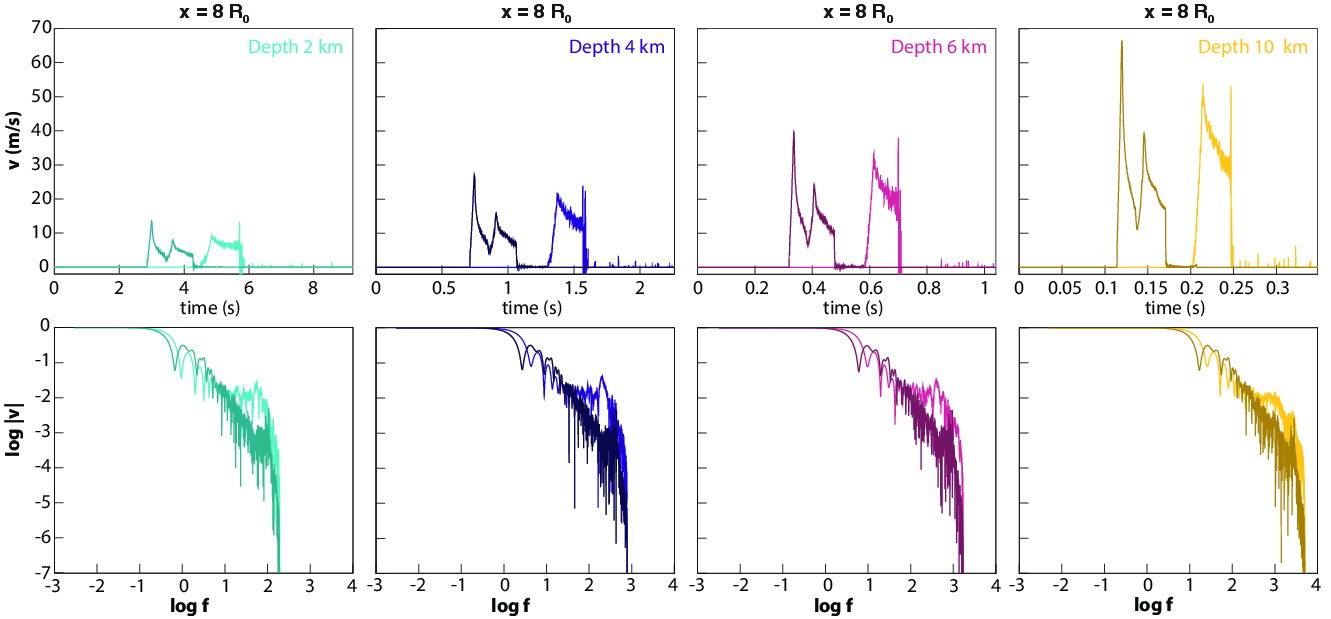}
    \caption{Comparison of Fourier amplitude spectra (FAS) of slip rate at different depths for the cases with an initial exponentially decreasing damage state. The dark and light colors correspond to the elastic and damage case respectively.}
    \label{fig:freq_analysis_expo}
\end{figure}

\section*{Spring-slider with normal stress and stiffness variations}

We have implemented the spring-slider model with varying normal stress described in \citeA{perfettini_frictional_2001}. This model enables us to replicate variations in load within the bulk, such as seasonal hydrological loading or tides. To simulate variations in bulk elastic properties, we incorporate varying spring stiffness into the aforementioned model.

\subsection*{Governing equations}

A spring-slider is a simple model where a rigid block connected to a spring of stiffness $k$ is pulled at a constant velocity $V_0$. Friction is governed by a rate-and-state friction which in \citeA{perfettini_frictional_2001} takes the following form \cite{ruina_slip_1983}:

\begin{equation}
	\tau = \sigma \left(\mu_0 +a \ln \left(\frac{V}{V_*}\right) + \psi \right),
    \label{eq:RSF_Perfettini}
\end{equation}

with $V_*$ a normalizing velocity and $\psi$ a state variable whose derivative is given by:

\begin{equation}
	\frac{d \psi}{d t} = - \frac{V}{D_c} \left(\psi + b \ln \left(\frac{V}{V_*}\right)\right) - \frac{\alpha}{\sigma} \frac{d \sigma}{d t}.
	\label{eq:dpsi}
\end{equation}
$\alpha$ is a normal stress variation coefficient between 0 and $\mu_0$ (see \citeA{perfettini_frictional_2001} for details). This rate-and-state formulation features a varying normal stress $\sigma(t)$ which is expressed as:

\begin{equation}
	\sigma(t) = \sigma_0 \left(1 + \epsilon f(t)\right), \qquad |\epsilon f(t)| \ll 1.
	\label{eq:BK_sigma}
\end{equation} 
The two forces at work are the spring restoring force and friction. Hence applying Newton's second law gives:

\begin{equation}
	m \frac{dV}{dt} = ku - \tau,
	\label{eq:Newton2}
\end{equation}
with $u$ the spring displacement and $\dot u = V_0 - V$. We approximate the system by saying $m = 0$ and add a radiation damping term $\eta V$. \cref{eq:Newton2} then becomes:

\begin{equation}
	\tau = ku - \eta V,
	\label{eq:tau2}
\end{equation}
with $\eta = \mu / (2 c_s)$ and $c_s$ the shear wave velocity. \cref{eq:RSF_Perfettini} and~\cref{eq:tau2} are equal. Differentiating them with respect to time one gets:

\begin{equation}
	\frac{dV}{dt} = \left[\frac{dk}{dt}u + k\left(V_0 - V \right) - \left(\mu_0 +a \ln \left(\frac{V}{V_*}\right) + \psi \right) \frac{d\sigma}{dt} -\sigma \frac{d \psi}{dt}\right] / \left[{\frac{a}{V}\sigma + \eta}\right].  
	\label{eq:dV}   
\end{equation} 
Finally, the system of ordinary differential equations is:
\begin{equation}
	\left\{
	\begin{array}{lll}
		\frac{du}{dt} = V_0 - V \\
		\frac{d \psi}{d t} = - \frac{V}{D_c} \left(\psi + b \ln \left(\frac{V}{V_*}\right)\right) - \frac{\alpha}{\sigma} \frac{d \sigma}{d t} \\
		\frac{dV}{dt} = \left[\frac{dk}{dt}u + k\left(V_0 - V \right) - \left(\mu_0 +a \ln \left(\frac{V}{V_*}\right) + \psi \right) \frac{d\sigma}{dt} -\sigma \frac{d \psi}{dt}\right] / \left[{\frac{a}{V}\sigma + \eta}\right].
	\end{array}
	\right.
	\label{eq:system}
\end{equation} 

\subsection*{Spring stiffness variations}

 Spring stiffness $k$ is the equivalent in our 1D system of the bulk elastic properties. Therefore, we use geologic and geodetic observations to set its evolution. A drop in seismic velocities followed by a partial recovery is observed after an earthquake \cite{froment_imaging_2014} and attributed to elastic properties changes due to damage. In our model, $k$ follows a law that reproduces this behaviour.
 Inspired by shape-similarity between seismic velocities evolution \cite{froment_imaging_2014} and friction coefficient evolution after a change in loading velocity, we define a velocity-dependent evolution law for $k$ similar to the rate-and-state law with a state variable $\theta_k$ for the spring. We thus have: 

\begin{equation}
	k(t) = k_0 \left(1 + b_k \ln \left(\frac{V_0 \theta_k}{D_k}\right)\right)
\end{equation} 
and:

\begin{equation}
	\frac{d \theta_k}{dt} = 1 - \frac{V \theta_k}{D_k},
	\label{eq:eq_k}
\end{equation}
with $k_0$ the initial spring stiffness. This differential equation is joined to \cref{eq:system} to form the governing system:

\begin{equation}
	\left\{
	\begin{array}{llll}
		\frac{du}{dt} = V_0 - V \\
		\frac{d \psi}{d t} = - \frac{V}{D_c} \left(\psi + b \ln \left(\frac{V}{V_*}\right)\right) - \frac{\alpha}{\sigma} \frac{d \sigma}{d t} \\
		\frac{dV}{dt} = \left[\frac{dk}{dt}u + k\left(V_0 - V \right) - \left(\mu_0 +a \ln \left(\frac{V}{V_*}\right) + \psi \right) \frac{d\sigma}{dt} -\sigma \frac{d \psi}{dt}\right] / \left[{\frac{a}{V}\sigma + \eta}\right] \\
		\frac{d \theta_k}{dt} = 1 - \frac{V \theta_k}{D_k}.
	\end{array}
	\right.
	\label{eq:system2}
\end{equation} 
This system is solved using a Bulirsch-Stoer method with adaptive time-stepping. It allows a fine temporal resolution when needed while staying computationally efficient. 

\subsection*{Settings}

 In this spring-slider model, the seismic velocity drop after an earthquake depends on the parameter $b_k$. Therefore, we set $b_k$ value to reproduce drop amounts observed by \citeA{froment_imaging_2014}.

 The initial spring stiffness $k_0$ is taken to be a fraction of the critical stiffness $k_c$ given by $k_c = \sigma_0 (b - a) / D_k$ as the system is unstable if $k < k_c$. Normal stress variations are sinusoidal, that is $f(t)$ in \cref{eq:BK_sigma} is equal to $\sin(2\pi t/T)$ with $T$ the period. All parameters used are summarized in \cref{table:param_BK}.

\begin{table}[h!]
	\centering
	\begin{tabular}{ |c|c|c| } 
		\hline
		\textbf{Symbol} & \textbf{Parameter} & \textbf{Value} \\ 
		\hline 
		$\mu$ & Shear modulus & 30 MPa \\ 
		$c_s$ & S-wave speed & 3500 m/s \\
		$\eta$ & Viscous term & $4.3 \times 10^{-6} \ \mathrm{kg.m^{-2}.s^{-1}}$ \\
		
		$\mu_0$ & Steady-state friction coefficient & 0.6 \\
		$a$ & R\&S friction parameter & 0.007 \\	
		$b$ & R\&S friction parameter & 0.009 \\	
		$V_*$ & Normalizing velocity & $10^{-4}$ m/s \\
		$D_c$ & Critical slip distance & 1.35 mm \\
		
		$\sigma_0$ & Mean normal stress & 100 MPa \\
		$\alpha$ & Normal stress variation coefficient & 0.1 \\
		$k_c$ & Critical spring stiffness & $14.8 \times 10^{7}$ N/m \\
		$k_0$ & Initial spring stiffness & $0.95 \times k_c$ \\
		
		$D_k$ & Characteristic healing distance & $5 \times 10^{-2}$ m \\	
		$T$ & Period of normal stress variation & 1 year \\	
		\hline
	\end{tabular}
	\caption{Parameters used for simulations with the spring-slider.}
	\label{table:param_BK}
\end{table}

\end{document}